\documentclass[aps,prl,twocolumn,showpacs,amsmath,amssymb]{revtex4}

\usepackage{graphicx}
\begin{document}

\newcommand{\UofA}{Department of Physics, University of Alberta, Edmonton, AB T6G 2G7, Canada}

\title{Spin wave dynamics and the determination of intrinsic damping
 in locally-excited Permalloy thin films}

\author{Zhigang Liu} \email{zgliu@Phys.UAlberta.CA}
\affiliation{\UofA}

\author{Fabian Giesen}
\affiliation{\UofA}

\author{Xiaobin Zhu}
\altaffiliation[Present address:]{~Seagate Research, 1251 Waterfront
Place, Pittsburgh, PA 15222}
\affiliation{\UofA}

\author{Richard D. Sydora}
\affiliation{\UofA}

\author{Mark R. Freeman}
\affiliation{\UofA}

\begin{abstract}

Time-resolved scanning Kerr effect microscopy has been used to study
magnetization dynamics in Permalloy thin films excited by transient
magnetic pulses generated by a micrometer-scale transmission line
structure. The results are consistent with magnetostatic spin wave
theory and are supported by micromagnetic simulations. Magnetostatic
volume and surface spin waves are measured for the same specimen
using different bias field orientations and can be accurately
calculated by $k$-space integrations over all excited plane wave
components.  A single damping constant of Gilbert form is sufficient
to describe both scenarios. The nonuniform pulsed field plays a key
role in the spin wave dynamics, with its Fourier transform serving
as a weighting function for the participating modes. The intrinsic
Gilbert damping parameter $\alpha$ is most conveniently measured
when the spin waves are effectively stationary.
\end{abstract}


\maketitle

Interest has been growing for many years in the time-domain
investigation of magnetization dynamics in response to a short
magnetic pulse. An impulse excitation is broadband, but the dynamics
of the magnetic system are also very sensitive to the parameters of
the excitation pulse. Time-domain pulse shaping can eliminate
ringing for a coherently-switched ferromagnetic element, to greatly
enhance the performance for applications
\cite{Gerrits2002,Schumacher2003}. In addition, the spatial profile
of the pulsed field is critical in dictating the magnetic dynamics.
Magnetostatic spin waves generated by such nonuniform transient
field have been observed in a number of time-resolved optical
\cite{Silva2002} and inductive \cite{{Covington2002,Wu2006}}
experiments, and also in frequency-domain studies \cite{Tamaru2004}.
The resulting position-dependent temporal response creates
additional challenges for characterizing the dynamics and for
determining the intrinsic magnetic damping. The focus of the present
work is to address these difficulties within a simple physical
framework.

\begin{figure} [t]
\centering
\includegraphics [height=3.375in,angle=270] {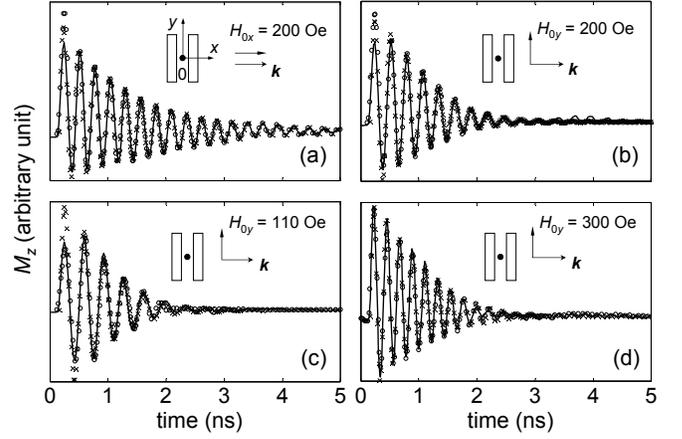}
\caption{\label{Fig1_Liu} Different damping behavior of local
magnetization under nonuniform excitation. The coordinate system is
defined in (a) and is the same throughout the work. The parallel
rectangular bars represent the stripline structure and the black
dots represent the probe points (not to scale). Relation between the
wave vector and the bias field is shown in each panel. The solid
curves are measured $M_z(t)$ traces, the crosses are calculated
results based on Eq.(\ref{Eq k-space integration}), and the open
circles are simulated results using the quasi-1D LLG model
\cite{early_mismatch}.}
\end{figure}

Under the condition of linear behavior of the spin wave dynamics in
the low amplitude regime \cite{{Covington2002,Wu2006}}, the magnetic
response is the linear superposition of all plane wave components
that can be excited by the pulse. Assuming the spin waves propagate
only along the $x$ direction (the coordinate system is defined in
Fig.\ref{Fig1_Liu}(a)), the out-of-plane component of magnetization
can be described by:
\begin{equation}
\
M_z(x,t)=e^{-\frac{t}{\tau}}\int_{0}^{k_c}P(k,\omega(k))\sin[kx-\omega(k)t+\phi]dk
\label{Eq k-space integration}
\end{equation}
where $\tau^{-1}$ is the inverse decay time, $\omega(k)$ is the
dispersion relation, and $P(k,\omega(k))$ is the spectral density of
the pulse field determined from its spatial and temporal profiles
and acts as a relative weighting factor for the different spin wave
components in $k$-space. The influences of the pulse field
parameters, the intrinsic damping, and the dispersion relation of
the spin waves are contained explicitly. $k_c$ is a cut-off wave
number for numerical integration ($k_c=5$ $\mu\textnormal{m}^{-1}$
is sufficient for the magnetostatic regime with the stripline
dimensions used here). The initial phase angle $\phi$ is taken to be
independent of frequency on account of the pulse excitation. The
longer trailing edge of the pulse causes a non-oscillatory
quasi-static background that is not included in Eq.(\ref{Eq k-space
integration}), but in actual calculations we take the pulse shape
into account by fitting the high-frequency data \cite{Elezzabi1996}.

The experimental geometry is shown by the inset schematics in
Fig.\ref{Fig1_Liu}. A $\textnormal{Ni}_{80}\textnormal{Fe}_{20}$
film (Permalloy, or ``Py''), with thickness $d=10\pm1$ nm, is
deposited on a 150 $\mu$m thick glass substrate using e-beam
evaporation. The film is then clamped on a copper thin film
stripline, with a small amount of high-vacuum dielectric grease
applied for a strong surface-tension bond and to ensure electrical
isolation between Py and Cu. Two coplanar wires in the stripline
structure transmit a current pulse (rise time $<20$ ps at the
sample) from a semi-insulating GaAs photoconductive switch (carrier
lifetime $\sim$ 300 ps) and generate a nonuniform magnetic pulse
$\textbf{\emph{h}}(x,t)$. The width and separation of the wires are
both 3 $\mu$m and are much smaller than the length ($\sim400$
$\mu$m), and the system can be treated as quasi-one-dimensional. The
pulse field amplitude (and corresponding initial torque) decreases
quickly away from the wires, falling to less than $10\%$ of the peak
value beyond $|x|=8$ $\mu$m for the experimental geometry. We define
the region of the Py film enclosed by these boundaries as the
``source'' area. An in-plane bias field $\textbf{\emph{H}}_0$
saturates the magnetization of the Py film such that
$\textbf{\emph{M}}_0$ is parallel to $\textbf{\emph{H}}_0$. This
layout fixes the directions of the wave vectors $\textbf{\emph{k}}$
to be parallel to $x$-axis, and enables us to detect different spin
wave modes by changing the direction of $\textbf{\emph{H}}_0$
($\textbf{\emph{M}}_0$). The focus here is on the special cases of
$\textbf{\emph{k}}\parallel{\textbf{\emph{M}}_0}$ and
$\textbf{\emph{k}}\perp{\textbf{\emph{M}}_0}$, although other angles
can be similarly addressed \cite{Giesen}. Changes of $M_z$ are
measured by means of time-resolved scanning Kerr effect microscopy
\cite{Hiebert1997}. This technique offers $\sim500$ nm spatial
resolution determined by the spot size of the focused probe beam
(much smaller than typical spin wave length encountered in this
work), and introduces a versatility relative to the pulsed inductive
method, whose spatial resolution is limited by fixed-position,
micrometer-size probe devices, and suffers a loss of signal when the
magnetization is perpendicular to the wires. The optical approach
allows a direct determination of a variety of spin wave dispersion
laws.

Quasi-1D micromagnetic modeling was also carried out in order to
benchmark the $k$-space calculation. The magnetic film was
discretized along the $x$-direction, such that the ``finite
elements'' were 10 nm in both $x$- and $z$-directions, while
infinitely long in $y$-direction. The spin motion of each element
obeys the Landau-Lifshitz-Gilbert (LLG) equation \cite{MiltatBook}:
\begin{equation}
\frac{d\textit{\textbf{M}}}{dt}=-\gamma_0\textit{\textbf{M}}\times\textit{\textbf{H}}_\textnormal{eff}
+\frac{\alpha}{\textit{M}_s}\textit{\textbf{M}}\times\frac{d\textit{\textbf{M}}}{dt}
\label{Eq LLG}
\end{equation}
where $\gamma_0$ = 17.61 MHz/Oe is the gyromagnetic ratio, $M_s$ =
760 emu/$\textnormal{cm}^3$ is the saturation magnetization of the
Permalloy film, and $\alpha$ is the Gilbert damping parameter.
$\textit{\textbf{H}}_\textnormal{eff}$ is the effective field mainly
contributed by the external field and the magnetostatic field. The
exchange interaction is found to be insignificant in the
magnetostatic regime \cite{Covington2002,Crawford2003}. The
implementation of the simulation follows standard procedures
\cite{{MiltatBook,Mansuripur1988}}.

Fig.\ref{Fig1_Liu}(a) shows the typical response of $M_z$, in the
$\textbf{\emph{k}}\parallel{\textbf{\emph{M}}_0}$ geometry, with
$H_{0x}=200$ Oe. The solid curve is a measurement at $x=0$, and
other positions show almost the same profile but with amplitude
decrease with increasing $x$. The damping is spatially uniform and
not influenced by the propagation of spin waves
\cite{Bailleul2003,Barman2004}, consistent with expectation for
$\textbf{\emph{k}}\parallel{\textbf{\emph{M}}_0}$. This geometry
exhibits magnetostatic backward volume waves with a dispersion law
\cite{Bailleul2001}:
\begin{equation}
\omega^{2}=\omega_H^2+\omega_H\omega_M(1-e^{-kd})/kd \label{Eq MSBVW
mode}
\end{equation}
where $\omega_H=\gamma_0H_0$, $\omega_M=4\pi\gamma_0M_s$. For the
present experiments, the group velocity, $v_g=\frac{d\omega}{dk}$,
is on the order of 0.1 $\mu$m/ns and the spin waves are effectively
stationary over the time scale of the measurement. The calculated
waveform based upon Eq.(\ref{Eq k-space integration}) is plotted by
the crosses in Fig.\ref{Fig1_Liu}(a), and agrees well with the
measured data. To determine the weighting function $P(k,\omega(k))$,
a standard linearization analysis is applied on Eq.(\ref{Eq LLG})
\cite{GurevichBook} to give
$M_z=\frac{\omega_M\omega_H}{\omega_H^2-\omega^2}h(\omega)h_z(k)$,
where $h(\omega)$ and $h_z(k)$ are the Fourier transformation of
$h(t)$ and $h_z(x)$, respectively. The $\omega$-dependent quantities
are found to be insignificant in the calculations (except for a
negative sign if $\omega>\omega_H$), so the weighting function can
be approximated by $P(k)=|h_z(k)|$. The Biot-Savart law was used to
calculate the in-plane ($h_x$) and out-of-plane ($h_z$) components
of the excitation field. The spatial distributions of $h_x$ and
$h_z$ depend on the distance $\Delta$ between the plane of the Py
film and the plane of the stripline. $\Delta$ cannot be precisely
measured here and is used as a fitting parameter. In this nearly
stationary case, magnetic damping of the system is unambiguously
determined by the exponential decay time $\tau$ in Eq.(\ref{Eq
k-space integration}), and can be measured directly from the
logarithm of the decreasing amplitude of the experimental waveform;
the result for Fig.\ref{Fig1_Liu}(a) is $\tau=1.40\pm0.01$ ns. On
the other hand, the Gilbert damping parameter $\alpha$ can be
independently fitted by micromagnetic simulations based solely on
LLG equation; the result for this case is $\alpha=0.0081\pm0.0003$,
and the simulated waveform is plotted by open circles in
Fig.\ref{Fig1_Liu}(a), in excellent correspondence with the
measurement and the $k$-space calculation. The two damping
parameters are related by $\tau=[\alpha\gamma_0 (2\pi
M_s+H_0)]^{-1}$ \cite{{GurevichBook}}, and our results obtained by
independent fittings are consistent with this relation.

Good agreement between the measurements, $k$-space calculations, and
quasi-1D simulations are again obtained in the
$\textbf{\emph{k}}\perp{\textbf{\emph{M}}_0}$ geometry, as shown in
Fig.\ref{Fig1_Liu}(b-d) for the probe positioned at $x=0$ and
$H_{0y}$ ranging from 110 Oe to 300 Oe. In this geometry
magnetostatic surface waves (MSSW) are the dominant modes, leading
to qualitatively different spatiotemporal dynamics in the Py film.
For the $k$-space calculations, the dispersion law of MSSW
\cite{{GurevichBook},{Sandler1999},{DE1961}},
\begin{equation}
\omega^{2}=(\omega_H+\omega_M/2)^2 -(\omega_M/2)^2\exp(-2kd)
\label{Eq DE mode}
\end{equation}
is used to calculate $M_z(x,t)$ using Eq.(\ref{Eq k-space
integration}). Linear response theory for the surface modes yields
$M_z=\frac{\omega_M}{\omega_H^2-\omega^2}h(\omega)[\omega_H
h_z(k)+i\omega h_x(k)]$, that is, the out-of-plane magnetization
responds to both in-plane and out-of-plane pulse fields. In the
present work $\omega_H\ll\omega$, the contribution of $h_z$ is small
and the weighting function again can be approximated by
$P(k)=|h_x(k)|$ \cite{CommentP(k)Shape} (the $\omega$-dependence is
neglected as in the
$\textbf{\emph{k}}\parallel{\textbf{\emph{M}}_0}$ case). The
$M_z(t)$ traces show significantly shortened decay time (compare
Fig.\ref{Fig1_Liu}(b) to Fig.\ref{Fig1_Liu}(a)), since the excited
surface modes possess fairly large group velocity to transfer the
nonequilibrium spin wave energy out of the probed position. At
$x=0$, the decay time shown in Fig.\ref{Fig1_Liu}(b-d) does not
change explicitly when $H_{0y}$ decreases (which leads to increasing
group velocity), but after we average $M_z(x,t)$ over
$|x|\leqslant8$ $\mu$m, the decay time in the whole ``source'' area
indeed decreases with larger group velocity, as expected (results
not shown). In other words, the damping behavior in MSSW
configuration cannot be quantitatively described by single-point
measurements, but has to be analyzed using the global approaches
(scanning probe experiment, micromagnetic simulation and $k$-space
calculation). The damping behavior is naturally embedded in
Eq.(\ref{Eq k-space integration}), because of dephasing of the
component frequencies through the $\omega(k)t$ term. Eq.(\ref{Eq
k-space integration}) is a generalization of the formula proposed in
Ref.\cite{Covington2002}, which gives an intuitive description of
the phenomenon. The damping envelope is jointly determined by a
Gaussian term accounting for spin wave dispersion, and the intrinsic
exponential decay. Our general approach works for both
$\textbf{\emph{k}}\parallel{\textbf{\emph{M}}_0}$ and
$\textbf{\emph{k}}\perp{\textbf{\emph{M}}_0}$ geometries, and
explains why the decay remains exponential when the spin wave
propagation is negligible.

\begin{figure} [t]
\includegraphics [width=3.375in,angle=270] {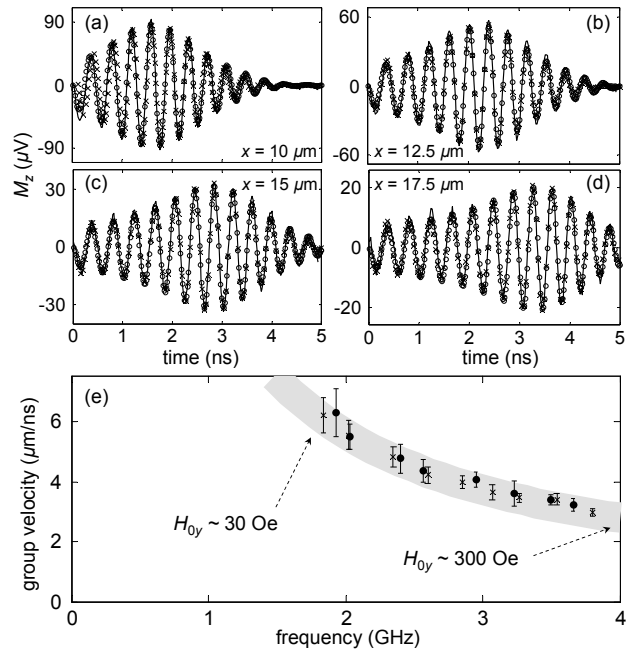}
\caption{\label{Fig2_Liu} Propagation of a spin wave packet in the
$\textbf{\emph{k}}\perp{\textbf{\emph{M}}_0}$ geometry. (a) -- (d):
$M_z(t)$ traces at $x$ = 10, 12.5, 15 and 17.5 $\mu$m, respectively,
when $H_{0y}=80$ Oe. The solid curves represent the measured data,
which are extracted from a single spatiotemporal scan, the crosses
are the calculated results based on Eq.(\ref{Eq k-space
integration}), and the open circles are the simulated results using
the quasi-1D LLG model. The simulated and calculated waveforms are
scaled to the same amplitude as the experimental data. (e): Group
velocity of MSSW modes as a function of oscillation frequency, with
$H_{0y}$ ranging from 30 to 300 Oe. The black dots are measured
data, and the crosses are results from quasi-1D simulations. The
grey shadow is the theoretical curve calculated from MSSW dispersion
law (Eq.(\ref{Eq DE mode})), with the width reflecting its lower and
upper limits due to the uncertainty in the cutoff wavelength and
film thickness ($k=0\sim5$ $\mu\textnormal{m}^{-1}$ and $d=9\sim11$
nm).}
\end{figure}

In the $\textbf{\emph{k}}\perp{\textbf{\emph{M}}_0}$ geometry,
individual wave packets propagating in the $x$ direction can be
observed when $M_z(t)$ is measured outside the ``source'' area
(i.e., the probe point moves away from the stripline).
Representative results for the case of $H_{0y}=80$ Oe are shown in
Fig.\ref{Fig2_Liu}(a-d). Recording a two dimensional (position and
time) map for the peak of the wave packet, its group velocity can be
determined to be $4.8\pm0.5$ $\mu$m/ns. Performing such analysis for
a range of bias fields yields the group velocity as a function of
frequency, as shown in Fig.\ref{Fig2_Liu}(e). The measurements and
numerical calculations agree reasonably well with the MSSW theory
($v_g$ determined from Eq.(\ref{Eq DE mode})).

For the propagating wave packet discussed above, the $M_z(t)$ is
asymmetric in time (the increase of the oscillation amplitude
appears ``slower'' than the following decline). This asymmetry is
especially apparent in Fig.\ref{Fig2_Liu}(c,d), but again is well
reproduced by the $k$-space calculation based on Eq.(\ref{Eq k-space
integration}) (and which cannot be achieved by the Gaussian-type
formula in Ref.[4]). This is shown by the crosses in
Fig.\ref{Fig2_Liu}(a-d), and is also supported by the quasi-1D
simulations (open circles). A single scale factor fits the measured
amplitude at all positions in the considered range, indicating that
the Gilbert mechanism ($\alpha=0.0081$ is still used here
\cite{CommentAlphaError}) also accounts for the spin wave
attenuation during propagation \cite{Covington2002,Bailleul2003}.
Measurements of this attenuation are potentially an effective way to
determine $\alpha$ when
$\textbf{\emph{k}}\perp{\textbf{\emph{M}}_0}$.

\begin{figure} [t]
\centering
\includegraphics [height=3.375in,angle=270] {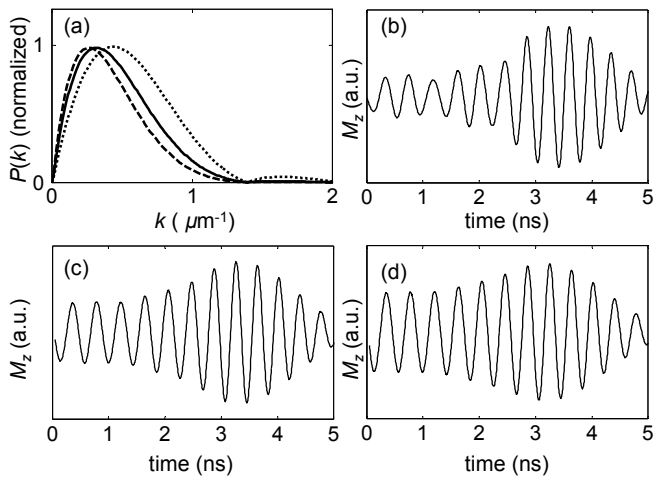}
\caption{\label{Fig3_Liu} Influence of the spatial distribution of
the pulse field. (a), Calculated distributions of normalized $P(k)$
for $\Delta=0.5$ $\mu$m (dotted curve), $\Delta=1.6$ $\mu$m (solid
curve), and $\Delta=2.5$ $\mu$m (dashed curve). (b) -- (d), $M_z(t)$
traces at $x=17.5$ $\mu$m calculated with Eq.(\ref{Eq k-space
integration}), using $\Delta=0.5$ $\mu$m, $\Delta=1.6$ $\mu$m, and
$\Delta=2.5$ $\mu$m, respectively.}
\end{figure}

The temporal profile of $M_z(t)$ ultimately stems from the spatial
profile of the excitation field, which directly determines $P(k)$ in
Eq.(\ref{Eq k-space integration}). Fig.\ref{Fig3_Liu} illustrates
the effect of different distributions $P(k)$, calculated for several
values of the film-stripline gap $\Delta$. As $\Delta$ increases,
the spatial variation of the pulse field becomes smoother and $P(k)$
acquires relatively higher spectral density at smaller $k$
(Fig.\ref{Fig3_Liu}(a)). This yields relatively larger oscillation
amplitude at early times before the wave packet peaks, and can be
understood here as a consequence of more spin wave components with
higher phase velocity $v_p=\frac{\omega}{k}$.
Figs.\ref{Fig3_Liu}(b-d) show the bracketing of the ``best-fit'',
$\Delta=1.6$ $\mu$m, to the experimental data as presented in
Fig.\ref{Fig2_Liu}(d).

\begin{figure} [t]
\centering
\includegraphics[height=3.375in,angle=90]{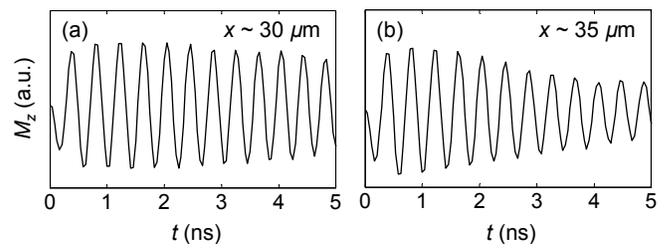}
\caption{\label{Fig4_Liu} Spin wave oscillations measured at (a),
$x\sim30$ $\mu$m and (b), $x\sim35$ $\mu$m. The experimental
conditions are the same as those in Fig.\ref{Fig2_Liu}(a-d).}
\end{figure}

As the spin wave packet propagates farther away from the source, its
shape is gradually broadened and the peak time eventually exceeds
the maximum optical delay of the apparatus (5 ns). In this
situation, the oscillations are dominated by small-$k$ components.
The experiments yield a unique position at each bias field
($x\sim30$ $\mu$m for the case of $H_{0y}=80$ Oe) where the incoming
energy from the propagating mode effectively balances the intrinsic
dissipation at that position, such that a nearly time-independent
oscillation amplitude is observed throughout the measurement window,
as shown in Fig.\ref{Fig4_Liu}(a).  For larger $x$, the power
balance is broken and the intrinsic decay of the long-wavelength
oscillations dominates (Fig.\ref{Fig4_Liu}(b)).

In summary, we have studied the spatiotemporal dynamics of a
ferromagnetic film in response to a short magnetic pulse localized
in one spatial dimension. The response can be interpreted as a
superposition of plane waves modulated by the spectral densities of
the spatially nonuniform, transient excitation field, and decaying
according to intrinsic Gilbert damping. Magnetostatic volume modes
and surface modes can be self-consistently described within this
interpretation, and the experimental and analytical results show
good agreement also with micromagnetic simulations. The Gilbert
damping parameter could be directly measured from broadband FMR
waveforms when the spin waves are effectively stationary. The
$k$-space calculations also require negligible computer resources in
comparison to micromagnetic simulations, and offer an opportunity to
invert the problem and design magnetic waveforms for applications.
In addition, the analysis can be extended to a second dimension to
account for thicker or multilayered structures, or for 2D spin wave
propagation.

This work was supported by NSERC, $i$CORE, CIAR and CRC. The samples
were fabricated in the University of Alberta Nanofab.

\begin{thebibliography} {99}
\bibitem{early_mismatch} The model pulse shape gives rise to a mismatch over the first cycle
of oscillation as measured at the center of the transmission lines.
This does not affect the conclusions concerning damping. The
propagating spin wave packets measured outside the transmission line
structure are insensitive to this detail of the pulse shape.
\bibitem{CommentAlphaError} Fitting to the data for surface waves is more complicated
than for volume waves due to stronger dispersion, and result in a
larger uncertainty. The fitting quality is in the same level for
$\alpha=0.0081\pm0.0006$.
\end{thebibliography}
\end{document}